# Fieldbus Device Drivers for Accelerator Control at DESY

H. G. Wu, DESY

Abstract

In order to interface the DESY fieldbus adapter, SEDAC (SErial Data Acquisition and Control system), a full duplex device driver was developed for the Windows NT, Linux, VxWorks, and Solaris operating systems. Detailed driver development issues as well as a common user interface will be presented, along with a comparison of the device drivers among the different operating systems. In particular, we shall present benchmark results concerning general performance as well as ease of development.

## 1 INTRODUCTION

The DESY in-house fieldbus adapter SEDAC has been used extensively at DESY since 1979 when it was designed. Thousands of hardware pieces, ranging from control magnets, BPMs to vacuum pumps are all connected with the SEDAC bus. SEDAC is a single master field bus. The SEDAC adapter (master) is via a 50 ohm BNC cable, over several kilometers, connecting up to 32 SEDAC slave crates. The individual hardware modules are residing in the slave crate. Data are transferred as 16 bits telegrams with 4 KHz rate, as read or write. There have been several generations of SEDAC adapters since then. In the past, during the upgrading of the machine control systems we have made several new SEDAC device drivers, which have a unified device access interface for using different versions of the adapters, and under various operating systems, such as Windows NT, Linux, VxWorks and Solaris.

## 2 HARDWARE COMPONENTS

One variant of the SEDAC adapters is a single slot ISA card with one SEDAC line. another one is a SEDAC IP (Industry Park) card with two lines, occupying two IP slots. For accessing the IP card we use two kinds of IP carrier card from Greenspring, ATC40 ISA bus adapter and PCI-40A PCI bus adapter for PC. This allows access 4 SEDAC lines through a single ISA or PCI slot of a PC. The VME162 CPU running VxWorks, with 4 on board IP slots, can also directly access 4 SEDAC lines. Both interrupt driven and polling mode of operation are possible.

Like other field buses the register structure of the SEDAC adapter is rather simple, consisting of READ/WRITE data, Status/Error and Command/Configuration registers. One can load data into the register, then write to the command register to start transmission. To determine telegram transmission completion one can either poll the status register or wait on interrupt.

## 3 DRIVER DEVELOPMENTS

### 3.1 Comparison of different operating systems

There are different approaches to implementing device drivers for various operating systems.

### 3.1.1 Windows NT

The kernel SEDAC device driver on Window NT is written with DDK (device driver kit) and SDK (software development kit). The development is rather easy because Window NT has a very powerful and rich I/O subsystem. Since the SEDAC adapter, and the IP to PCI or ISA adapters are rather old, requiring hardware jumpers, the SEDAC device driver under Windows NT is still a legacy driver, without a Plug and Play feature such as WDM (Windows Driver Model) for Window 2000.

There are several important features for the kernel SEDAC driver:

- It is object based design, the kernel objects are extensively used, such as *I*/O Request Packets (IRPs), Driver Objects, Device Objects and Device Extension, Interrupts Objects, Device Queue Objects and Deferred Procedure Call (DPC) Objects.
- There is one kernel device queue (FIFO) for each SEDAC IO device (line), the block of telegrams of all connected SEDAC lines can be parallel transmitted.
- It handles multiple adapters simultaneously.
- It provides synchronous and asynchronous calls.
- A kernel timer keeps track of the passage of time during transmission for detecting timeout.
- A cancel I/O routine is attached to each SEDAC IRP, the waiting telegrams submitted with asynchronous calls can be canceled by the user program. For an improperly

- terminated program it is critical to remove these telegrams away.
- An event logger is installed to log important events, such as device errors, into the System Event Log Buffer, for debugger purpose.
- DPC is used for cleaning up packets, which makes the driver more responding.
- It can request and release hardware resources from/to the system via the registry to avoid conflict with other drivers.

Several utility calls are provided, for giving status of the driver to applications. The diver can be called to find out how many programs has connected to the driver, how many telegrams waiting in the queue, and a call to cancel its own submitted telegrams.

We tested the kernel driver in polling mode, by adding either a kernel thread or a timer function. Since the minimum time interval in NT kernel is 10 ms, which is much longer then SEDAC telegram interval ~250us, the thread or timer based polling driver is not suitable.

It is easy to port the driver to WINDOW 2000 without many changes.

### 3.1.2 Linux

A similar kernel mode driver for Linux (kernel 2.0 and 2.2) is written. Most of the coding was directly copied from the driver on Window NT. Although the environment of Window NT and Linux operating systems are very different, the handling of data transfer and control operations for the adapter are the same.

Both blocking and non-blocking calls, as well as a select() mechanism are implemented.

The asynchronous mode is not support directly by the driver. A separate user thread, with shared memory can be used as a middle layer for making asynchronous calls.

The driver offers direct bus IO memory map, mmap() call, to application. For running full SEDAC bandwidth, the application can map the SEDAC register IO memory into user space, then use polling mode to transmit telegrams.

The driver registration is down via command lines during driver loading, by specifying I/O address and interrupt numbers.

### 3.1.3 VxWorks

We use VxWorks OS with SEDAC for the HERA magnet system. The front-end machine is a SUN workstation running Solarias with 2 CPUs, the VxWorks memory, as a slave, can be seen from the SUN via the SBUS. In such a way, the SEDAC telegram queues (a circular buffer) are mapped both for VxWorks and Solaris. In order to have full bandwidth, we choose polling mode for SEDAC device access. The driver is operating in the user space, as a single task owned device driver. A simple data structure, submitted from front-end server (SUN), to the queue is specified, the VxWorks has a dedicated task, which is responding for polling the queue at 4 KHz rate. After it wakes up, it examines all the SEDAC lines and checks whether there are waiting telegrams queued by the SUN. If there are any, it loads the telegrams into the adapter, then waits for a response via semaphores. The interrupt service routine copies data and status, then releases the semaphore for further data transmission.

### 3.2 Driver Interface

The same user interfaces are used both for Windows NT and Linux.

In the case of a driver running under Windows NT, we developed a DLL and an OCX (OLE Control eXtension) ActiveX control, for accessing the driver. One can use direct DLL calls to read or write block of telegrams synchronously. For asynchronous call one can use sedac.ocx. When transmission completes, the SEDAC completion event will be fired. For Linux a similar interface routine was written. An application can open a SEDAC line, then use simple read or write calls to transmit telegrams.

The interface routine, requested from an open line call from an application, will open the SEDAC character device file (/dev/sedacX). Then the application can fill in a data buffer, which contains the SEDAC line number, crate address and sub address, to read or write. In response to a read or write call, the interface routines DeviceIoCOntrol() or ioctl() are used to access the device driver.

### 3.3 Benchmarks

The overhead of the kernel driver depends heavily on the CPU clock rates the number of connected SEDAC lines. For a 133 Mhz Pentinum PC running Window NT, the overhead is about 80us, while about 25us for running Linux 2.2. This time is proportional to the number of SEDAC lines, which are simultaneously in operation. If one use block telegrams, the transfer overhead is negligible, because the data transfers are all handled down at interrupt level without lots of context switches between application and kernel driver. The large difference of the driver overhead between running Window NT and Linux shows that the NT system is handling many more resources, particularly as it cannot be run in a console mode (i.e. no WINDOWS) as can Linux.

The driver on VxWorks, as a single user mode driver, is operating on full bandwidth of 4 KHz.

## CONCLUSION

There are similarities for writing a kernel device driver for different operating systems. If one driver is written, the others can be quickly ported with minor changes.

The single user device driver in user space is most efficient and fast, as in the case of SEDAC device driver for HERA magnet system running VxWorks .

The SEDAC device driver has shown all the advantages of a kernel device driver. It has high performance when interrupt driven. It is easy to add new features and adding more devices. It is also capable of serving multitask operation, and handling synchronous requests from multithreads.